\documentstyle[12pt,aps,eqsecnum,amsmath,epsfig]{revtex}
\tightenlines
\begin{document}

\title{Coherence correlations in the dissipative two-state system}
\author{
  Gunther Lang $^{1}$,   Elisabetta Paladino $^{2,1}$, 
  and Ulrich Weiss $^{1}$   }
 \address{ 
 $^{1}$ Institut f\"{u}r Theoretische Physik, Universit\"{a}t
           Stuttgart, 70550 Stuttgart, Germany\\
  $^{2}$   Istituto di Fisica, Universit\`a di Catania {\rm\&} INFM,
           Viale A. Doria 6, 95129 Catania, Italy.}
\maketitle

\begin{abstract}

We study the dynamical equilibrium 
correlation function of the polaron-dressed tunneling operator 
in the dissipative two-state system. Unlike the position operator,
this coherence operator  acts in the full system-plus-reservoir space.
We calculate the relevant modified influence functional and 
present the exact formal expression for the coherence correlations
in the form of a series in the number of tunneling events.
For an Ohmic spectral density with the particular damping strength 
$K=\frac{1}{2}$, the series is summed in analytic form for all times
and  for arbitrary values of  temperature and bias. Using a diagrammatic 
approach, we find the long-time dynamics in the
regime $K<1$. In general, the coherence correlations decay algebraically 
as $t^{-2K}_{}$  at $T=0$. This implies that the linear static susceptibility 
diverges for $K\le\frac{1}{2}$ as $T\to 0$, whereas it stays finite for 
$K>\frac{1}{2}$ in this limit. The qualitative differences with respect  to 
the asymptotic behavior of the position correlations  are explained.

\end{abstract}


\section{Introduction}
The simplest model that allows to study the interplay of tunneling and 
dissipation is the spin-boson model~\cite{leggett,book}. 
Despite its simplicity, it exhibits generic features of many complex 
systems and has found widespread applications in physics and chemistry.
 It has been adopted to describe diverse systems, like the tunneling of atoms
between a surface and the tip of an atomic-force microscope \cite{micr}, 
or the dynamics of the trapped flux in a quantum interference 
device \cite{squid}, to mention a few.

For Ohmic dissipation, the spin-boson model shows a transition between 
coherent and incoherent tunneling where the location of the transition depends
on the damping strength and on the bias \cite{book}. 
Most investigations have been done for the non-equilibrium
expectation value $\langle \sigma_z(t) \rangle $,
representing the population difference between the two localized states, and 
for the $\sigma_z$ autocorrelation function, describing position or population
correlations \cite{sass-w901,sass-w902}. For the latter quantity,
the analogy with the Kondo model and the $1/r^2$ Ising model has been 
utilized~\cite{chakra,costi-ki,voelker} in numerical computations.
For Ohmic damping and zero temperature and bias, 
$\langle \sigma_z(t) \rangle$ shows a
transition between damped oscillations and incoherent relaxation
exactly at $K=\frac{1}{2}$~\cite{egg-grab-w}.
Recently, it has been argued~\cite{lesage97} that the quality factor 
of the oscillation is exactly given by $Q=\cot[\pi K /2(1-K)]$, which 
again gives $K=\frac{1}{2}$ for the transition.
For the
antisymmetrized $\sigma_z$ autocorrelation function, the same quality factor
and thus the same transition point was found  numerically~\cite{voelker}.
This is not in contrast to the earlier result 
$K=\frac{1}{3}$~\cite{costi-ki,les-sal-skor}, since there a different
criterion for the transition was applied~\cite{voelker,stockb}.

It has  been shown that the particular initial preparation plays a crucial
role for the long-time behavior at zero temperature. The factorized
system-reservoir initial state for the expectation value 
$\langle\sigma_z^{}(t)\rangle$ leads to
exponential decay \cite{egg-grab-w,lesage97}, whereas the symmetrized
$\sigma_z^{}$ equilibrium correlation function with a correlated
initial state decays algebraically as $1/t^2_{}$ for $K<1$,
as shown for the spin-boson model \cite{sass-w902} and the related 
$1/r^2_{}$ Ising and fermionic models \cite{chakra,costi-ki,les-sal-skor}. 
The power $2$ in the algebraic decay law is a signature of Ohmic dissipation.

Recently, focus has been put on expectation values \cite{GWW} and
equilibrium correlations \cite{strong} connected with the
tunneling  operator $\sigma_x^{}$.
The expectation value $\langle \sigma_x (t)\rangle $  and
the equilibrium  autocorrelation function of $\sigma_x$ have been found to be 
nonuniversal, i.e., vanish in the scaling limit~\cite{GWW,guinea}.
Here, we study the equilibrium autocorrelation function of a
polaron-dressed tunneling operator $\tilde\sigma_x$ which includes the 
adiabatic dynamics of the bath modes \cite{leggett,silb-har}. This
function is universal and measures correlations of the off-diagonal 
elements (coherences) of the density matrix.
We present the exact formal solution for the coherence correlations 
in the form of a series in the number of tunneling events. We then analyze
the resulting expression in various limits. In particular, we work out
the differences in the asymp\-totic decay between the position correlations
and the coherence correlations. Our analytical real-time approach
is complementary to the recent imaginary-time numerical studies in 
Ref.~\cite{strong}.

In Section II, we formulate the problem and introduce the correlation 
function $C_x(t)$ of the coherence operator $\tilde\sigma_x$.   
Since this operator acts both in the system and bath space, the elimination 
of the bath modes has to be reconsidered. The relevant considerations leading 
to a modified influence functional are given in Section III.
These results are used in Section IV to determine exact formal 
expressions for $C_x(t)$.
In Section V, we present the analytical solution for $C_x(t)$ for the special
value  $K=\frac{1}{2}$. Section VI is devoted to the regime 
$K=\frac{1}{2}-\kappa$ with $\kappa\ll 1$.
Finally, in Section VII we show that the asymptotic decay of 
$C_x(t)$ at zero temperature is algebraic with a
$K$-dependent power for $0<K<1$.

\section{Formulation of the problem}

It has been well established that the dissipative dynamics of a particle 
in a double-well potential can effectively be described at very low $T$
by the spin-boson model~\cite{leggett,book}
\begin{align}
\label{H2}
H &= H_0 + \sum_{\alpha} \bigg[\,\frac{p{}_\alpha^2}{2m_\alpha} + \frac{1}{2} 
        m_\alpha \omega_\alpha^2 \left(x_\alpha 
        - \frac{c_\alpha}{m_\alpha \omega{}_\alpha^2} \frac{a}{2} 
          \sigma_z \right)^2 \bigg] \;, \\  \label{Hspin}
H_0 &= - \hbar \left( \Delta \sigma_x + \epsilon \sigma_z \right)/2\; .
\end{align}
Here the basis is formed by the two localized states $|R\rangle$ and 
$|L\rangle$ 
of the double-well which are eigenstates of $\sigma_z$ with eigenvalues
$+1$ and $-1$, respectively. The position operator is $q= a \sigma_z /2$
 with $\sigma_z=|R\rangle\langle R|-|L\rangle\langle L|$.   
The tunneling operator $\sigma_x=|R\rangle\langle L|+|L\rangle\langle R|$
transfers the particle between the two wells with tunneling amplitude
$\Delta$. The second term in Eq.~(\ref{Hspin}) describes an externally 
applied bias energy $\epsilon$.
The effect of the thermal bath on the system's dynamics is included in the
spectral density 
\begin{equation}
\label{spectr.d.}
J(\omega) = \frac{\pi}{2} \sum_{\alpha} \frac{c{}_\alpha^2}
             {m_\alpha \omega_\alpha}
           \,\delta(\omega-\omega_\alpha) \;.
\end{equation}
The important case of an Ohmic bath is described by
\begin{equation}
\label{ohmicd.}
J(\omega) = \eta \, \omega \, e^{-\omega / \omega_c} = 
            ( 2 \pi \hbar  K / a^2)\,\omega\,e^{- \omega / \omega_c} \;,
\end{equation}
where $\eta$ is the viscosity, $K$ is the appropriate dimensionless 
damping strength, and $\omega_c$ is a cut-off for the bath modes.
We are interested in the regime  $\Delta\ll\omega_c$, in
which $\Delta$ and $\omega_c$ form a renormalized 
frequency scale~\cite{leggett}
\begin{equation}
\label{deltared}
\Delta_r = \Delta ( \Delta/\omega_c ) ^ {K/(1-K)} \;.
\end{equation}
A quantity is called universal, if it is a function of
 $\Delta_r$ alone, i.e., there is
no other $\omega_c$ dependence than given by Eq.~(\ref{deltared}).
Vice-versa, any extra dependence on $\omega_c$ is non-universal:
 performing the scaling limit 
$\Delta_r/\omega_c\to 0$ with $\Delta_r$ fixed, this contribution 
vanishes. 
Both the mean value of $\sigma_z$ and its equilibrium
autocorrelation function are universal~\cite{leggett,sass-w901}.
On the other hand, the expectation value   $\langle \sigma_x(t) \rangle$ is
equipped with an overall factor 
$\Delta_r / \Delta = (\Delta_r / \omega_c)^{K}$ and therefore vanishes in the
scaling  limit ~\cite{GWW}.

Here we concentrate on the $\sigma_x$ equilibrium autocorrelation function. 
As observed in Ref.~\cite{guinea} and explained in the sequel, the 
equilibrium correlation function of the bare $\sigma_x$
does not satisfy the above universality criterion. 
To overcome this shortcoming, we consider a modified tunneling operator
which takes into account the adiabatic displacement of the bath modes
during the tunneling process.
The transformation to a basis of displaced harmonic oscillators states 
is accomplished by the polaron unitary transformation~\cite{leggett}
\begin{equation}
\begin{split}
U &= \exp \left \{ - i \sigma_z \Omega / {2 \hbar} \right \} \; \\ 
\label{omega}
\Omega &=a\sum_{\alpha}\frac{c_\alpha}{m_\alpha\omega_\alpha^2}p_\alpha
       =  \sum_{\alpha} s_\alpha p_\alpha \;.
\end{split}
\end{equation}
The set of displacements is given by
$\{s_\alpha \} \equiv \{ a c_\alpha/m_\alpha \omega{}_\alpha^2 \}$.
The polaron transformed tunneling operator 
$\tilde{\sigma}_x =  U \sigma_x U^{-1}$ reads
\begin{align}\label{sigma2}
        \tilde{\sigma}_x 
        & =  |R\rangle\langle L| \exp \{ -i \Omega/\hbar \} +
                      |L\rangle \langle R|\exp \{i \Omega/\hbar \}\\ 
\label{sigmac}& = |R \rangle   \langle L| \int
             dx \,|x \rangle \, \langle x-s|\, +\; \mbox{h.\, c.}\;\;,
\end{align}
where we have introduced the compact notation
\begin{equation}
s \equiv \{s_\alpha \} \;,\quad x \equiv \{ x_\alpha \}\;,\quad
\int dx  \equiv 
                       \prod_{\alpha} \int dx_\alpha
\;.
\end{equation}
The bare $\sigma_x$ acts in the Hilbert space of the two-state system (TSS) 
alone, whereas the dressed operator $\tilde{\sigma}_x$ acts in the
full system-plus-reservoir space.
From the coordinate representation (\ref{sigmac}), we immediately see that 
the operation of $\tilde{\sigma}_x$  transfers the particle from 
one localized state to the other and simultaneously shifts 
 each bath-oscillator by the 
displacement $\pm s_\alpha$ (``polaronic cloud''). 
In this sense, $\tilde{\sigma}_x$ generates coherent tunneling between the 
two localized states and can be called coherence operator
\cite{coherent}. The coherence correlations are then described by
the equilibrium correlation function of  $ \tilde{\sigma}_x$,
\begin{equation}
\label{fcorr1}
C_x (t) = \langle \tilde{\sigma}_x (t) \tilde{\sigma}_x (0) \rangle _\beta 
        = \mbox{Tr}\,[\, \tilde{\sigma}_x (t) 
            \tilde{\sigma}_x (0) W_\beta \,]\;,
\end{equation}
where $W_\beta=e^{-\beta H} /\,\mbox{Tr} \,[\, e^{-\beta H}\,]$ is the
 equilibrium density matrix of the global system,
and $ \tilde{\sigma}_x(t)$ is the  Heisenberg-representation of 
$ \tilde{\sigma}_x$ with respect to the untransformed 
Hamiltonian (\ref{H2}) (cf. Ref.~\cite{transform}).
The associated response function $\chi_x(t)=(-2/\hbar)\,\Theta(t)\,{\rm Im}
\,C_x(t)$ describes the linear response of the system to a
coherence inducing perturbation $H_{\rm pert}\propto \tilde\sigma_x$.

It is  convenient to consider
$C_x(t)$ as the mean value of  $\tilde{\sigma}_x(t)$
with respect to the ``density matrix'' $W=\tilde{\sigma}_x(0)W_\beta$.
 Switching to 
the Schr\"{o}dinger-picture, Eq.~(\ref{fcorr1}) becomes 
\begin{equation} 
\label{fcorr2}
C_x (t) =\mbox{Tr} \, [ \, \tilde{\sigma}_x  W(t) \,] \;,  
\end{equation}
where the time-dependent ``density matrix''
$W(t) = \exp (- i H t / \hbar) \, W(0) \, \exp ( i H t / \hbar)$
obeys the initial condition $W(0) = \tilde{\sigma}_x  W_\beta$.
Inserting the expression (\ref{sigmac}) for $\tilde{\sigma}_x$ into 
Eq.~(\ref{fcorr2}) 
and performing the trace, we find that $C_x(t)$ is the sum of the 
off-diagonal matrix elements 
\begin{equation}
\label{c.f.2}
C_x(t) = \, \rho{}_{1, -1}^{(s)} (t) \, + \, \rho{}_{-1, 1}^{(-s)} (t) \;.
\end{equation}
Since  $\tilde{\sigma}_x$ acts also in the bath space, 
$\rho{}_{i,j}^{(s)}(t)$ is different from the usual reduced density 
matrix as it appears, e.g., in the 
$\sigma_z$ correlation function~\cite{sass-w901}. 
We shall refer to $\rho{}_{i,j}^{(s)}(t)$ as the 
``shifted reduced density matrix'' (SRDM).
To be general, we now give the discussion for a continuous variable $q$, 
and return to the two-state system only in Section IV.
We have 
\begin{equation}\label{sdm2}
\begin{split}
\rho^{(s)}(q_f , q_f',t) &= \int dx_f \>
                      \langle q_f, x_f + s|\, W(t) \,| q_f', x_f \rangle\\ 
&=\int dq_i\,dq_i'\,dx_f\,dx_i\,dx_i'\,K(q_f,x_f+s,t;q_i,x_i,0)\\
& \quad\times K^*(q_f',x_f,t;q_i',x_i',0)\langle q_i, x_i |\,W(0)\,|q_i', 
x_i' \rangle\;,
\end{split}
\end{equation}
where $K(q_f,x_f+s,t;q_i,x_i,0)$ is the usual Feynman propagator that may be 
expressed as a path-integral. The matrix elements of the initial 
``density matrix'' $W(0)=\tilde\sigma_x W_{\beta}$ read 
\begin{equation}
\label{in.cond.}
\langle q_i, x_i | \, W(0) \, | q_i', x_i' \rangle =
 \langle -q_i, x_i-s\,\mbox{sgn}(q_i) | \, W_\beta \, | q_i', x_i' \rangle\;, 
\end{equation}
where the bath coordinates are also affected by the preparation.
Now, it remains to integrate out the shifted bath degrees of freedom in the 
expression 
(\ref{sdm2}).

\section{General initial preparations}

The standard Feynman-Vernon approach which may be used to eliminate the
bath degrees of freedom relies on the assumption of a factorized 
system-bath initial state. For ergodic systems, it is possible to 
obtain a real-time description also for a thermal initial state ~\cite{book}. 
This approach can be generalized to special classes of correlated initial 
states by introducing a preparation function~\cite{grab-schr-ing}. 
For  the case of $C_x(t)$, the method in Ref.~\cite{grab-schr-ing}
 has to be reconsidered, since the 
initial preparation also involves the bath. 
To proceed, we define a generalized preparation function 
$\lambda^{}_{\rm{G}} ( q_i, q{}_i';\,\bar{q}, \bar{q}';
 x_i, x{}_i';\,\bar{x}, \bar{x}')$ by
\begin{equation}
\label{prepgen}
\langle q_i, x_i |\, W(0)\, | q{}_i', x{}_i' \rangle  =
  \int \!\! d\bar{q} \; d\bar{q}' \;  d\bar{x} \;  d\bar{x}' \;
   \lambda^{}_{\rm{G}} ( q_i, q{}_i'; \,\bar{q}, \bar{q}';
             x_i, x{}_i';\,\bar{x}, \bar{x}') \;  
\langle \bar{q}, \bar{x} |\, W_\beta\, | \bar{q}', \bar{x}' \rangle\;.
\end{equation}
Comparing this form with Eq.~(\ref{in.cond.}),
we see that the preparation function factorizes as 
\begin{equation}
\label{factprepf}
\lambda^{}_{\rm{G}} ( q_i, q{}_i'; \,\bar{q}, \bar{q}';
             x_i, x{}_i';\,\bar{x}, \bar{x}') =
\lambda^{}_{\rm{S}} ( q_i, q{}_i'; \,\bar{q}, \bar{q}') 
\lambda^{}_{\rm{R}} ( x_i, x{}_i'; \,\bar{x}, \bar{x}')\;, 
\end{equation}  
where the system's and reservoir's preparation functions are given by
\begin{align}
\label{prepq}
   \lambda^{}_{\rm{S}} ( q_i, q{}_i'; \,\bar{q}, \bar{q}') 
   &= \delta ( q_i +  \bar{q} )
    \; \delta ( q{}_i'  -  \bar{q}' )\;, \\
\label{prepx}
\lambda^{}_{\rm{R}} ( x_i, x{}_i';   \bar{x}, \bar{x}') 
   &= \delta ( x_i - \bar{x}   -  s \, \mbox{sgn}(q_i))
    \; \delta ( x{}_i'  -  \bar{x}' ) \;. 
\end{align}
With the form (\ref{factprepf}), the evolution of the SRDM (\ref{sdm2}) is 
given by
\begin{align}
\label{rhos}
\rho^{(s)}( q_f, q{}_f', t )&=\int dq_i\,dq{}_i'\,d\bar{q}\,d\bar{q}'\,
   J^{}_{\rm{G}}(q_f, q{}_f',t; q_i,q{}_i'; \bar{q}, \bar{q}')
\lambda^{}_{\rm{S}} ( q_i, q{}_i'; \,\bar{q}, \bar{q}') \;,
\end{align}
where the generalized propagating function reads
\begin{align}\label{prop}
 J^{}_{\rm{G}}(q_f, q{}_f',t; q_i,q{}_i'; \bar{q}, \bar{q}') 
     &=
   \int dx_f  \, dx_i \, dx{}_i' \, d\bar{x} \, d\bar{x}' \,
   K(q_f, x_f+s, t; q_i, x_i, 0) \notag \\
   &\quad\times\,
   \langle\bar{q},\bar{x}|\,W_\beta\,|\bar{q}',\bar{x}'\rangle \;
K^*(q_f, x_f, t;  q_i, x_i, 0) \;
\lambda^{}_{\rm{R}}(x_i, x{}_i'; \bar{x}, \bar{x}')
\;.
\end{align}
For an ergodic system, the thermal density matrix $W_\beta$
in Eq.~(\ref{prop}) can be expressed as follows. 
Describe the global system at a time $t_0 < 0$ by a factorized density matrix, 
the system being in a position eigenstate, say $|q_0\rangle$, and the 
reservoir being in thermal equilibrium, $W(t_0)= |q_0\rangle\langle q_0|
\otimes\, e^{-\beta H_{\rm{R}}} / \mbox{Tr}\,
 [ \, e^{-\beta H_{\rm{R}}} \,]$~\cite{book}, and let it
evolve out of this state under the full Hamiltonian. Then,
 if $t_0$ is sent to 
the infinite past, the system has reached at time zero the correlated 
initial state $\langle\bar{q},\bar{x}|\,W_\beta\,|\bar{q}',\bar{x}'\rangle$.
With these considerations, we may rewrite (\ref{prop}) as
\begin{align}
\label{genprop}
J^{}_{\rm{G}}(q_f, q{}_f',t; q_i,q{}_i',0^+&; \bar{q}, \bar{q}',0^-;
 q_0,q_0',t_0) \notag\\
   & =\int\! {\cal D} q \int\! {\cal D} q'
         \exp \left [ \frac{i}{\hbar} 
                     \left ( {\cal S}^{}_{\rm{S}} [q] 
                         - {\cal S}^{}_{\rm{S}} [q'] \right ) 
              \right] {\cal F}^{}_{\rm{G}} [ q, q'; s]\;,
\end{align}
where ${\cal S}^{}_{\rm{S}}[q]$ is the action corresponding to the system
Hamiltonian (\ref{Hspin}). The functional integrations are over all paths 
$q(t')$ and $q'(t')$ which satisfy the constraints
\begin{align} 
\label{constrainq1}
 q(t_0)&=q_0 & q(0^{-})& = \bar{q} &  q(0^{+})&=q_i & q(t)&=q_f \\
\label{constrainq2}
 q'(t_0)&=q{}_0'&  q'(0^{-})&= \bar{q}'& q'(0^{+})&=q{}_i' &q'(t)&=q{}_f'
\; .
\end{align}
All the effects of the bath onto the system's dynamics are captured by the 
generalized influence functional
\begin{align}
\label{inflg}
  {\cal F}^{}_{\rm{G}} [ q, q';s] &=
  \int \! dx_f \; dx_i \; dx{}_i' \;  d\bar{x} \; d\bar{x}' 
     \;  dx_0  \; dx{}_0' \; W^{}_{\rm{R}}(x_0, x{}_0') \;
  \lambda^{}_{\rm{R}} ( x_i, x{}_i';\bar{x}, \bar{x}')   \notag \\
  &\quad\times\,\int\! {\cal D} x \int\! {\cal D} x'
         \exp \left [ \frac{i}{\hbar} 
                     \left ( 
                            {\cal S}^{}_{\rm{R, I}} [x, q] 
                            - {\cal S}^{}_{\rm{R, I}} [x',q'] 
                     \right ) 
                \right]\;,
\end{align} 
where ${\cal S}_{\rm{R, I}} [x, q]$ is the action corresponding to the 
reservoir and interaction terms in Eq.~(\ref{H2}). 
The paths $x(t')$ and $x'(t')$ are subject to the constraints
\begin{align} 
\label{constrainx1}
 x(t_0)&=x_0 & x(0^{-})& = \bar{x} &  x(0^{+})&=x_i & x(t)&=x_f+s \\
\label{constrainx2}
 x'(t_0)&=x{}_0'&  x'(0^{-})&= \bar{x}'& x'(0^{+})&=x{}_i &x'(t)&=x{}_f\;.
\end{align}
Compared to the standard Feynman-Vernon influence functional, there are
 two differences. First, the endpoint of the $x$-path is shifted by 
the displacement $s$, i.e., the bath does not end up in a diagonal 
state at time $t$. 
Second, the reservoir paths $x(t')$ and $x'(t')$ are discontinous at 
time zero, depending on the reservoir's preparation function 
$\lambda^{}_{\rm{R}} ( x_i, x{}_i';\bar{x}, \bar{x}')$.

\section{Exact formal solution}

Having obtained an explicit expression for the SRDM
at time $t$ for a general initial preparation, we can now write down
the exact formal solution for the coherence correlations $C_x(t)$, 
Eq.~(\ref{c.f.2}).  
Inserting the preparation function (\ref{prepq}) into the SRDM
(\ref{rhos}), we get
\begin{align}
\label{c}
C_x(t) &=\lim_{t_0\to -\infty}\sum_{q_i, q{}_i', q_f}
     J^{}_{\rm{G}}(q_f, \, -q_f, \, t; \, q_i, \, q{}_i',0^+;
        \,  -q_i, \,  q{}_i',0^-;q_0,q_0,t_0)\;,\notag
\end{align}
where $q_i,\; q_i',\; q_f=\pm a/2$.
Note that there is a jump in the $q-$path at time zero.
Inserting the reservoir preparation function (\ref{prepx}) into the
influence functional (\ref{inflg}), the
 $x'(t')$-path (\ref{constrainx2}) turns out  continuous at time $t'=0$,
 whereas the $x(t')$-path  (\ref{constrainx1})
is discontinuous. Because of the integration over $x_i$, the constraints
 (\ref{constrainx1}) may equivalently be expressed as  
\begin{align}
 x(t_0)&=x_0 & x(0^{-})& = x_i &  x(0^{+})&=x_i+s\,\mbox{sgn}(q_i) 
             & x(t)&=x_f+s\,\mbox{sgn}(q_f)\;. 
\end{align}

Consider  first the  contributions to $C_x(t)$ with $q_i=q_f$. In this case,
the shifts in the $x$-path at times $t'=0^{+}$ and $t'=t$ are 
equal, and thus we can eliminate the  shift at positive times by defining 
 modified reservoir coordinates  according to 
\begin{equation}\label{xmod}
\tilde{x}(t') = x(t') - s\, \mbox{sgn}(q_i) \Theta(t')\;. 
\end{equation}
The path $\tilde{x}(t')$ is  continuous at $t'=0$ and obeys
 $\tilde{x}(t)=x_f$. In the shifted coordinate, the bath is in a
 diagonal state at time $t$. 
As the  action $S^{}_{\rm{R,I}}[x,q]$ appearing in the influence 
functional is quadratic both in $x(t')$ and in $q(t')$ and bilinear 
in the coupling, the second term in Eq. (\ref{xmod}) can be absorbed  
into a modified  $q$-path which is continuous at $t'=0$, 
\begin{equation}
  \tilde{q}(t') = q(t') - 2 q_i \Theta(t')\;.
\end{equation}
Writing the influence functional  in terms of 
the paths $q'(t'),\; x'(t')$ and the modified paths  
$\tilde{q}(t'),\;  \tilde{x}(t')$, the displacement $s$ is completely
 eliminated from the description. Thus, after integrating out the bath 
degrees of freedom, we end up with an influence functional that is of 
the standard Feynman-Vernon form for a factorized initial state at 
time $t_0$,
\begin{equation} \label{influenza}
{\cal F}^{}_{\rm{G}} [ q, q'; s \, \mbox{sgn}(q_i)] = 
   {\cal F}[\tilde{q}, q']\;.
\end{equation}
All effects in $\tilde{\sigma}_x$ induced by the  polaronic cloud 
are in the modified path $\tilde{q}(t')$. 

Next, consider the contributions to (\ref{c}) with $q_i=-q_f$. Now, it is
 not possible to end up with an influence functional of the form 
(\ref{influenza}) in which the shifts of the bath modes are fully
 absorbed into a modified path $\tilde{q}(t')$. In the usual charge picture 
(see below), the case $q_i=-q_f$ corresponds to sequences of
 charges that violate overall neutrality. As a result, $\Delta$ and 
$\omega_c$ cannot be combined to a function of $\Delta_r$ alone. 
Instead, each contribution comes with an additional factor 
$(\Delta_r/\omega_c)^{4K}$ and therefore is nonuniversal. Thus in the 
scaling limit $\Delta_r/\omega_c\rightarrow 0$, all contributions
 with $q_i=-q_f$ vanish. 

\begin{figure}[htb]
\setlength{\unitlength}{0.85cm}
\centerline{
\begin{picture}(9,4.75)
\put(0,2.05){\epsfig{file=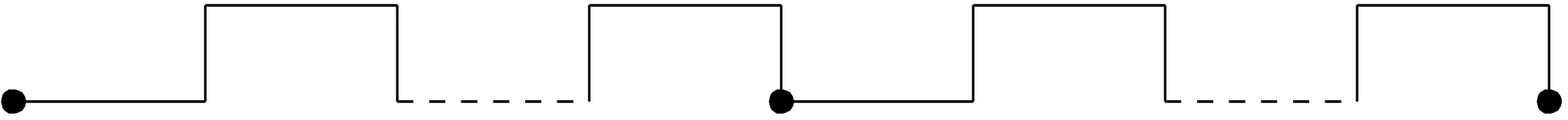,width=0.572cm,angle=-90}}
\put(0,0.7){$t_0$}
\put(3.9,0.7){$t_{2n}=0$}
\put(7.5,0.7){$t_{2n+2m}=t$}
\put(0,3.08){$t_0$}
\put(1.6,1.6){$\xi_1$}
\put(3.8,1.6){$\xi_n$}
\put(5.85,1.6){$\xi_{n+1}$}
\put(8.02,1.6){$\xi_{n+m}$}
\put(3.9,3.08){$t_{2n+1}=0$}
\put(7.5,3.08){$t_{2n+2m}=t$}
\put(1.6,4){$\xi_1$}
\put(4.72,4){$\xi_{n+1}$}
\put(8.02,4){$\xi_{n+m}$}
\put(0,4.43){\epsfig{file=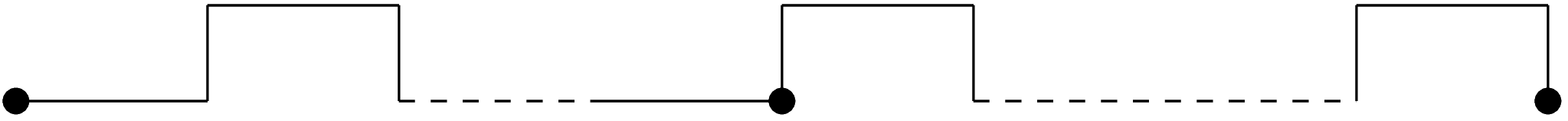,width=0.583cm,angle=-90}}
\end{picture}}
\caption{Path contributions to group A (top) and group B (bottom). The 
steps represent blips of either sign, and sojourns are indicated by the
 baseline.\label{generic} }
\end{figure}

With the above, the correlation function is now given by 
\begin{equation}
\label{cx1}
C_x(t) = \lim_{t_0 \to -\infty}\sum_{q_i, q{}_i'}
     J^{}_{\rm{G}}(q_i, \, -q_i, \, t; \, q_i, \, q{}_i',0^+;
        \,  -q_i, \,  q{}_i',0^-;q_0,q_0,t_0)\;.
\end{equation} 
At this stage, it is important to note that the free propagators in the
 propagating function 
depend on the original paths $q(t')$ and $q'(t')$. The concept of
  modified paths is only used to express the generalized influence 
functional in the standard Feynman-Vernon form.
For the evaluation of Eq.~(\ref{cx1}), it is convenient to introduce the
linear combinations 
\begin{equation}
\begin{split} 
\eta(t)&=\big[\,q(t)+q'(t)\,\big]/a\;,\\
\xi(t)&=\big[\,q(t)-q'(t)\,\big]/a \;,
\end{split}
\end{equation}
describing propagation along the diagonal of the density matrix and
 off-diagonal excursions, respectively.
For the two-state system, these paths are piecewise constant with 
jumps at times $t_j$. As usual, the time 
intervals $t_{2j}<t'<t_{2j+1}$ in which the system is in  a diagonal state 
are called $sojourns$, while the time
 intervals  $t_{2j-1}<t'<t_{2j}$ spent in an off-diagonal state are
referred to as $blips$. A  sojourn is labelled by $\eta_j=\pm 1$, 
depending on whether the system is in state $RR$ or $LL$. 
Similarly, $\xi_j=\pm 1$ describes a blip  in which the system is in 
state $RL$ or $LR$. The lengths of the sojourn- and blip-intervals are 
denoted by $s_j= t_{2j+1}-t_{2j}$ 
and $\tau_j= t_{2j}-t_{2j-1}$, respectively. All paths that contribute
 to the correlation function (\ref{cx1}) start out from the initial 
 sojourn $\eta_0$ at time $t_0$ and end in the blip state  $\xi$
 at time $t$. According to their behavior at time zero, they can be divided  
into two groups. In  group A, the 
system jumps at time zero from a sojourn to a blip state ($q_i'=-q_i$ at
time $0^+$). In group B, the system hops at time zero from a blip to a
sojourn state ($q_i'=q_i$ at time $0^+$).
A general path with $n$ blips at negative and $m$ blips at positive times 
can  be parametrized by
\begin{equation}\label{chixi}
\begin{split}
\eta(t')&=\sum_{j=0}^{n+m-1}\eta_j\big[\,\Theta(t'-t_{2j})-  
\Theta(t'-t_{2j+1})\,\big] \;, \\
\xi(t')&=\sum_{j=1}^{n+m}\xi_j\big[\,\Theta(t'-t_{2j-1})-  
\Theta(t'-t_{2j})\,\big]\;,
\end{split}
\end{equation}
with $t_{2n+2m}=t$~\cite{charget}. For group A, we have  $t_{2n+1}=0$, 
whereas $t_{2n}=0$ for group B. According to the boundary conditions, 
the paths are subject to  the constraints 
\begin{align}\label{cnstrA}
\xi_{n+m} &=\xi_{n+1}=-\eta_n & &\mbox{(group A)}\;, \\
\xi_{n+m} &=-\xi_{n}=\eta_n   &   &\mbox{(group B)}\label{cnstrB}\;.
\end{align}
Generic contributions to group A and group B are sketched in 
Fig.~\ref{generic}.

Thus the correlation function is built up by two parts that correspond
 to these two different path classes. We have 
$C_x(t)=C_x^{\rm{A}}(t)+C_x^{\rm{B}}(t)$ with 
\begin{equation}\label{ca}
\begin{split}
C_x^{\rm{A}}(t) &=\lim_{t_0 
\to -\infty} \sum_{\xi}J^{}_{\rm{G}}(\xi,t;\xi,0^+;\eta=-\xi,0^-;
\eta_0,t_0 )\; ,\\ 
C_x^{\rm{B}}(t) &=\lim_{t_0 \to -\infty}
\sum_{\xi}J^{}_{\rm{G}}(\xi,t;\eta=\xi,0^+;-\xi,0^-;\eta_0,t_0) \; .
\end{split}
\end{equation}
The path sum is over all sequences of blips and sojourns  and implies  
time-ordered integration over the jump times. We introduce  the compact 
notation
\begin{align}\label{timeint}
  \int_{t_0}^t{\mathcal D}_{k,l}\{t_j\}&=
       \int_0^t\! dt_{k+l+1}\!\!\int_0^{t_{k+l+1}}\!\!
dt_{k+l}\,
           \ldots
       \int_0^{t_{k+3}}\!dt_{k+2}\int_{t_0}^0 dt_{k}\,
            \ldots
       \int_{t_0}^{t_2}\! dt_{1}\;.
\end{align}
Here  $k$ and $l$ represent the number of  flips in the time regimes
$t_0<t'<0$ and $0<t'<t$, respectively.  For group A, we have 
$k=2n,\, l=2m-2$,
 whereas  for group B  $k=2n-1$ and $l=2m-1$. Each  transition in 
Eq.~(\ref{timeint}) comes with a factor $\pm i\Delta/2$. There are  two
 additional transitions at time zero, $t_{k+1}=0$,
 and at time $t$,  $t_{k+l+2}=t$. These two hops, however, come without a
 factor $\pm i\Delta/2$  since they  are not dynamical. The jump at time
zero is enforced by the operation of $\tilde{\sigma}_x$,
whereas the jump at time $t$ is introduced for convenience 
(cf.~Ref.~\cite{charget}).  The amplitude to stay in a 
sojourn is unity, while the amplitude to stay in  blip $\xi_j$ is given by 
$\exp(i\epsilon\xi_j\tau_j)$. Thus, a full  path gives for both groups a
 factor  
\begin{align}
%  \mathcal{A}[q]\mathcal{A}^*[q']&\rightarrow
-\eta_0\,\xi(-\Delta^2/4)^{n+m-1}D_{n,m}
\end{align}
with the bias term 
\begin{align}
  D_{n,m}&=\exp\Big(i\epsilon\sum_{j=1}^{n+m} \xi_j \tau_j\Big)\;.
\end{align}

Before discussing the modifications due to the polaron transformation,
 consider the standard influence functional. Performing
 integrations by parts, it takes the form~\cite{book}
\begin{align} 
\label{uinfl}
{\mathcal F}[\eta,\xi]&=\exp\biggl\{\int_{t_0}^{t^+}\! dt'\!
                                  \int_{t_0}^{t'}\!dt''\,
\Big[\,\dot{\xi}(t')S(t'-t'')\dot{\xi}(t'')
 +i\dot{\xi}(t')R(t'-t'')\dot{\eta}(t'')\,\Big]
\biggr\} \;,
\end{align}
where the  kernels $S(t)$ and $R(t)$  are the real and imaginary part
 of the second integral of the bath correlation function. In the limit 
 $\omega_c t\gg 1$, we have~\cite{book}
\begin{align}\label{s}
S(t)&=2K\ln\big[(\hbar\beta\omega_c/\pi)\,
                 \mbox{sinh}(\pi t/\hbar\beta)\big]\;, \\
R(t)&=\pi K\, \mbox{sgn}(t)\label{r}\;.
\end{align}
Because of the form (\ref{chixi}), the velocities in 
Eq.~(\ref{uinfl}) consist of a series of delta-functions centered at 
the flip times. This suggests to regard the blip- and sojourn-paths as 
 sequences of charges: blip-charges interact with each other through the
 kernel $S(t)$, while the sojourn-charges interact with the  blip-charges 
via $R(t)$. Substituting the paths (\ref{chixi}) into Eq.~(\ref{uinfl}), 
the influence functional takes the form
\begin{equation}
\label{Ffv}
{\mathcal F}_{n,m} = G_{n,m}H_{n,m}\;.
\end{equation}
The factor  $G_{n,m}$ contains all the inter- and intrablip-interactions, 
\begin{align}
\label{G}
  G_{n,m} &= \exp \Big[ - \sum_{\substack{j=1}}^{n+m} S_{2j, 2j-1} -
                  \sum_{\substack{j=2}}^{n+m} \sum_{\substack{k=1}}^{j-1}
                  \xi_j \xi_k \Lambda_{j,k} \Big]\;, \\  \label{Lambda}
\Lambda_{j,k} &= S_{2j,2k-1} + S_{2j-1,2k} - S_{2j,2k} - S_{2j-1,2k-1} \;,
\end{align}
where $S_{p,q}=S(t_p-t_q)$.
The  sojourn-blip interactions are captured by the phase factor $H_{n,m}$.
 With the form (\ref{r}), each sojourn only interacts with the 
subsequent blip,
\begin{equation}
\label{H}
   H_{n,m} = \exp\Big[i\pi K\sum_{\substack{k=0}}^{n+m-1}
\eta_k\xi_{k+1}\Big]\;.
\end{equation}

Substituting Eq.~(\ref{s}) into the term $\Delta^{2n+2m}G_{n,m}$, 
the quantities 
$\Delta$ and $\omega_c$ are combined into a factor $\Delta_r^{(2-2K)(n+m)}$,
where $\Delta_r$ is the renormalized tunneling frequency, 
Eq.~(\ref{deltared}). The autocorrelation function of the bare $\sigma_x$ 
depends on the standard influence functional (\ref{Ffv})--(\ref{H}).
In this case, however, there appears the quantity 
$\Delta^{2n+2m-2}G_{n,m}$ because
the $\Delta$-factors of the  two blip-charges at time zero and time $t$ 
are missing. Therefore the autocorrelation function of the bare 
$\sigma_x$ comes with an overall factor
$\Delta_r^2/\Delta^2=(\Delta_r/\omega_c)^{2K}$. Thus it is nonuniversal
in the sense discussed above.

Now return to the correlation function $C_x(t)$ which
depends on the generalized influence functional ${\mathcal F}^{}_{\rm{G}}$. 
As shown 
above, this can be expressed in the standard form (\ref{uinfl}) if we
 substitute the modified paths
\begin{equation}\label{chiximod} 
\begin{split}
  \tilde{\xi}(t')&=\xi(t')-\xi \big[\,\Theta(t')-\Theta(t'-t)\,\big]\\
 \tilde{\eta}(t')&=\eta(t')-\xi \big[\,\Theta(t')-\Theta(t'-t)\,\big]\;.
\end{split}
\end{equation}
  The effects of the subtractions in Eq.~(\ref{chiximod})  are directly 
seen in the charge
 pic\-ture. Taking into account the constraints (\ref{cnstrA}) and 
(\ref{cnstrB}), one gets the following changes:
In the path $\tilde{\xi}(t')$, the two blip-charges at times $t'=0$
 and $t'=t$ are cancelled. 
In the path $\tilde{\eta}(t')$, the sojourn-charge originally located
 at time $t'=0$ is moved to time $t'=t$. 
It turns out that the influence functionals for the paths of group A
and group B are different. We write
\begin{align}
{\mathcal F}_{n,m}^{\rm{A}}&=G^{\rm{A}}_{n,m}H^{\rm{A}}_{n,m} \;,&
{\mathcal F}^{\rm{B}}_{n,m}&=G^{\rm{B}}_{n,m}H^{\rm{B}}_{n,m}\;.
\end{align}
The blip-interaction-factors $G^{\rm {A/B}}_{n,m}$ differ from the 
standard $G_{n,m}$ by the absence of the two blip charges at $t'=0$ and 
$t'=t$. For group A, this is
\begin{equation}
\label{GA}
  G^{\rm{A}}_{n,m} = \exp \Big[ - \sum_{\substack{j=1\\j\neq n+1}}^{n+m} 
                S_{2j, 2j-1} -
             \sum_{\substack{j=2}}^{n+m} \sum_{\substack{k=1}}^{j-1}
               \xi_j \xi_k \Lambda^{\rm{A}}_{j,k} \Big] \;,
\end{equation}
where $\Lambda^{\rm{A}}_{j,k}$ describes the interblip correlations for the
 modified
 sequence of charges. If $j,\,k\neq n+1$ and $\neq n+m$, 
$\Lambda^{\rm{A}}_{j,k}$ is
 again given by  (\ref{Lambda}). In all other cases, the interactions of the
 missing charges have to be dropped in (\ref{Lambda}). For instance, for
 $j=n+1$, we have $\Lambda^{\rm{A}}_{n+1,k}=S_{2n+2,2k-1}-S_{2n+2,2k}$.
 Similarly, we obtain for group B
\begin{equation}
\label{GB}
  G^{\rm{B}}_{n,m} =\exp\Big[-\sum_{\substack{j=1\\j\neq n}}^{n+m} 
S_{2j, 2j-1} -
            \sum_{\substack{j=2}}^{n+m} \sum_{\substack{k=1}}^{j-1} 
             \xi_j \xi_k \Lambda^{\rm{B}}_{j,k} \Big] \;,
\end{equation}
with analogous modifications in $\Lambda^{\rm{B}}_{j,k}$ for $j,\,k=n$ 
and $n+m$.
For instance, we have $\Lambda^{\rm{B}}_{n,k}=S_{2n-1,2k}-S_{2n-1,2k-1}$.  
The modified phase factors $H^i_{n,m}$ take the form
\begin{equation}\label{HB}
\begin{split}
   H^{\rm{A}}_{n,m}&=\exp\Big[i\pi K\sum_{\substack{k=0\\k\neq n}}^{n+m-1}
              \eta_k\xi_{k+1}\Big] \;, \\ 
   H^{\rm{B}}_{n,m}&= \exp\Big[i\pi K\sum_{\substack{k=0\\k\neq n}}^{n+m-1} 
            \Big(\eta_k\xi_{k+1}+\eta_n(\xi_{n+1}-\xi_{n+m}) \Big) \Big]\;.
\end{split}
\end{equation}
Thus each sojourn interacts with the subsequent blip except for sojourn $n$.
For group A, the sojourn $n$ is effectively noninteracting, whereas 
 for group B, it effectively interacts both with  blip $n$ and blip $n+m$. 

At this point, let us briefly reflect what we have gained so far.
First of all, since the sequence of the remaining $2n+2m-2$ blip charges is
neutral and comes with  a factor $\Delta^{2(n+m-1)}$, the quantities
$\Delta$ and $\omega_c$ are combined into a factor $\Delta_r^{(2-2K)(n+m-1)}$.
Thus, the  $\tilde{\sigma}_x$ autocorrelation function turns out to 
be universal. 
There is, however, an essential difference between the two groups.
For group A, the charges in the negative and positive time branches are 
neutral individually. For group B, there is an excess charge $\pm 1$ in 
each branch, and only the  combined arrangement is neutral again.
Since the asymptotic decay of equilibrium correlation functions crucially
depends on the interactions between the negative and positive time branches,
we should expect different behaviors for group A and group B.

Collecting the various results, we obtain explicit expressions
for the propagating functions of group A and group B in Eq.(\ref{ca}),
\begin{align}
  J^{}_{\rm{G}}&(\xi,t;\xi,0^+;\eta=-\xi,0^-;\eta_0,t_0)  \notag\\
     &=-\eta_0\xi\sum_{n=0}^{\infty}\sum_{m=1}^{\infty}
          \left(-\frac{\Delta^2}{4}\right)^{n+m-1}
     \int_{t_0}^{t}\! {\mathcal D}_{2n,2m-2}\{t_j\}
           \sum_{\{\xi_j\}^{\rm{A}}}G_{n,m}^{\rm{A}} D_{n,m} 
\sum_{\{\eta_j\}^{\rm{A}}}H_{n,m}^{\rm{A}}\;,
\end{align}
\begin{align}
   J^{}_{\rm{G}}&(\xi,t;\eta=\xi,0^+;-\xi,0^-;\eta_0,t_0)   \notag\\
     &=-\eta_0\xi\sum_{n=1}^{\infty}\sum_{m=1}^{\infty}
          \left(-\frac{\Delta^2}{4}\right)^{n+m-1}
         \int_{t_0}^{t}\! {\mathcal D}_{2n-1,2m-1}\{t_j\}
            \sum_{\{\xi_j\}^{\rm{B}}}G_{n,m}^{\rm{B}} D_{n,m}
 \sum_{\{\eta_j\}^{\rm{B}}}H_{n,m}^{\rm{B}}\;.
\end{align}
The summation is over all $\xi_j,\,\eta_j=\pm 1$. The superscripts 
$\{ \ldots \}^{\rm{A}}$
and  $\{ \ldots \}^{\rm{B}}$ indicate the constraints (\ref{cnstrA})
and (\ref{cnstrB}) with $\xi_{n+m}=\xi$, respectively. Using Eqs.~(\ref{ca}), 
$C_x(t)$ is obtained. It is now straightforward to perform the 
$\eta$-summations and to use symmetry
relations under exchange $\{\xi_j\}\rightarrow\{-\xi_j\}$. Taking the
limit $t_0\rightarrow -\infty$, the correlation function becomes independent 
of the initial value $\eta_0$.  In the end, we find  
for the symmetrized correlation function 
$S_x(t)=\mbox{Re}\,C_x(t)$ and the response function
$\chi_x(t)=(-2/\hbar)\,\Theta(t)\,\mbox{Im}\,C_x(t)$ the expressions

\begin{align}
S_x^{\rm{A}}(t)  &= \frac{1}{2}\sum_{m=1}^{\infty} 
                 \left(-\bar{\Delta}^2\right )^{m-1}
          \int_{-\infty}^{t}\!{\mathcal D}_{0,2m-2}\{t_j\}
            \sum_{\{\xi_j\}^{\rm{A}}}
                G_{0,m}^{\rm{A}} D^{(+)}_{0,m}  \label{sxa} \; ,\\
 S_x^{\rm{B}}(t) &= -\sum_{n=1}^{\infty}\sum_{m=2}^{\infty}
             \left(-\bar{\Delta}^2\right )^{n+m-1} \sin^2(\pi K)
             \int_{-\infty}^{t}\! {\mathcal D}_{2n-1,2m-1}\{t_j\}\
 \!\!\!\!\!\!\!\sum_{\substack{\{\xi_j\}\\\!\!\!\!\!\!\!\!\!\!\!\!\!
                       \xi_n=\xi_{n+1}=-\xi_{n+m}}} 
       \xi_1\,\xi_{n+m}\, G_{n,m}^{\rm{B}} D^{(+)}_{n,m} \;, \label{sxb}\\
\chi_x^{\rm{A}}(t)   &= \frac{1}{\hbar}\sum_{n=1}^{\infty}\sum_{m=1}^{\infty} 
             \left(-\bar{\Delta}^2\right )^{n+m-1}
               \tan(\pi K)\int_{-\infty}^{t}\! {\mathcal D}_{2n,2m-2}
            \{t_j\}
             \sum_{\{\xi_j\}^{\rm{A}}} 
            \xi_1\,\xi_{n+m}\,G_{n,m}^{\rm{A}}D^{(+)}_{n,m}\;,\label{axa}\\
\chi_x^{\rm{B}}(t) &= \frac{1}{\hbar}\sum_{n=1}^{\infty}\sum_{m=1}^{\infty} 
              \left(-\bar{\Delta}^2\right )^{n+m-1}
                  \tan(\pi K) \int_{-\infty}^{t} \! 
                   {\mathcal D}_{2n-1,2m-1}\{t_j\}   \notag  \\ 
            &\qquad \times \sum_{\{\xi_j\}^{\rm{B}}} 
               \xi_1 \, G_{n,m}^{\rm{B}} D^{(+)}_{n,m} 
            \;\{\sin^2(\pi K) \, \xi_{n+1} + \cos^2(\pi K) \, \xi_{n+m} \}\;. 
\label{axb}
\end{align}
Here we have introduced $ \bar{\Delta}^2=\Delta^2\cos(\pi K)/2$ and
\begin{align}
\label{D+}
   D^{(+)}_{n,m}&=\cos\Big(\epsilon\sum_{j=1}^{n+m} \xi_j \tau_j\Big)\;.
\end{align}
Equations (\ref{sxa})--(\ref{D+}) are exact formal series expansions for the
symmetrized equilibrium correlation function $S_x(t)$ and the response
function $\chi_x(t)$. Despite their formidable appearance, we can obtain
exact results in certain limits. This is discussed in the remainder of this
work. For the subsequent analysis, it is convenient to 
switch from integrations over the flip times, Eq.~(\ref{timeint}),
to integrations over blip lengths $\tau_j^{}$ and sojourn lengths $s_j^{}$. 

\section{The case $ K=\frac{1}{2}$}

For the value $K=\frac{1}{2}$, the above series for $S_x(t)$ and
 $\chi_x(t)$
can be summed in analytical form using the concept of collapsed blips and
collapsed sojourns \cite{sass-w901}. 
Putting $K=\frac{1}{2}-\kappa$ with $\kappa \ll 1$, the phase 
factor  
$\cos(\pi K)\approx \pi\kappa$ vanishes in the limit 
$\kappa\rightarrow 0$.
In order to have a finite contribution for $K=\frac{1}{2}$, each factor 
$\cos(\pi K)$
has to be compensated by a $1/\kappa$ singularity arising from the
``short-distance'' singularity of the breathing mode integral
of a dipole (blip or sojourn) with interaction 
$ e^{-S(\tau)} \approx (\omega_c \tau)^{-(1-2\kappa)}$. Thus we have 
\begin{equation}
\label{gamma}
I(K={\textstyle \frac{1}{2}})=
\lim_{K\rightarrow  1/2}\,\Delta^2 \cos(\pi K)
     \int_0 d\tau \, e^{-S(\tau)} \, 
=\frac{\pi}{2}\frac{\Delta^2}{\omega_c} \equiv \gamma  \;.
\end{equation}
We shall refer to an expression of the form (\ref{gamma}) 
as a collapsed dipole.
Since a collapsed dipole has zero dipole moment, it does not interact with
other charges. Further, it is insensitive to a symmetric bias factor.
In contrast, an odd bias factor in Eq.~(\ref{gamma}) prevents a 
dipole from collapsing, and combined with a factor 
$\cos(\pi K)$, this term  vanishes as $K\to \frac{1}{2}$.

A blip or a sojourn becomes extended when the $\cos(\pi K)$ factor
and the short-distance sin\-gularity are absent.
Within an extended blip of length $\tau$, 
the system may make any number of 
visits of duration zero to a sojourn and then returns to the same
blip. Mathematically, this is described
by the insertion of a grand-canonical ensemble of noninteracting collapsed 
so\-journs (CS), yielding a CS form factor $e^{-\gamma \tau/2}_{}$.
Likewise, within a sojourn of length $s$, the system may make
any number of visits of duration zero to a blip state. This is
represented by a grand-canonical ensemble of noninteracting collapsed 
blips (CB). Since the system can return to either sojourn state, there is a 
multiplicity factor 2, yielding a CB form factor $e^{-\gamma s}_{}$.   

An extended sojourn, say $s_k$, remains free of insertions only if the
subsequent blip is weighted with an unconstrained factor $\xi_{k+1}$.
In this case,  the $\{\xi_j\}$ summation leads to cancellations among the
interactions stretching over the extended sojourn, and thus it remains bare.
It turns out that this is a general rule also for $K\neq \frac{1}{2}$,
referred to as the $\xi$-rule in the sequel.
For the correlation functions 
 (\ref{sxb})--(\ref{axb}), e.g.,  the initial sojourn  
starting at $t_0$ remains bare  due to the factor $\xi_1$
in the exact formal expressions. There are no other bare intervals in the 
negative-time branch. Thus, the limit $t_0\to-\infty$ is
well-behaved.

\begin{figure}[htb]
\setlength{\unitlength}{0.75cm}
\centerline{
\begin{picture}(9,2.9)
\put(0,0.7){$t_0$}
\put(4.5,0.7){$0$}
\put(9,0.7){$t$}
\put(0,2){\epsfig{file=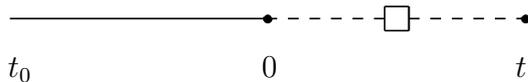,width=0.381cm,angle=-90}}
\end{picture}}
\caption{\small The diagram describing $S_x(t)$. The full and dashed lines 
represent sojourns and blips, respectively. The empty box represents the
insertion of a CS form factor within the blip interval. 
The bullets mark transitions which are free of bath correlations 
because of the modified influence functional.  
\label{figureS12}}
\end{figure}

Based on these concepts, we now analyze the various contributions.
Consider first the symmetrized correlation function. 
Assigning the $\cos(\pi K)$ factors
to the collapsing dipoles as in Eq.~(\ref{gamma}), there is one $\cos(\pi K)$
factor left in Eq.~(\ref{sxb}). Thus the contribution $S_x^{\rm{B}}(t)$
 vanishes  linearly with $\kappa$ as $K\to\frac{1}{2}$.
In $S_x^{\rm{A}}(t)$, the system dwells in the initial sojourn state $\eta$
 until time zero. At this time, it hops into the blip state $\xi=-\eta$
where it stays until 
time $t$, resulting in a factor $\cos(\epsilon t)$. The blip of length $t$ is
decorated with a CS form factor. Piecing the various components together, we 
find the damped oscillatory behavior
\begin{equation} \label{sxa12}
S_x(t)=S_x^{\rm{A}}(t)=\cos(\epsilon t)\,e^{-\gamma t/2}\;.
\end{equation}
The contributions to $S_x^{\rm{A}}(t)$ are sketched diagrammatically in 
Fig.~\ref{figureS12}.
Since only collapsed sojourns contribute to $S_x^{}(t)$ and the short-distance
behavior of the pair interaction is independent of temperature, the expression
(\ref{sxa12}) is valid at any temperature.

\begin{figure}[htb]
\setlength{\unitlength}{0.75cm}
\centerline{
\begin{picture}(9,7.3)
\put(0,0.7){$t_0$}
\put(4.5,0.7){$0$}
\put(9,0.7){$t$}
\put(0,3.7){\epsfig{file=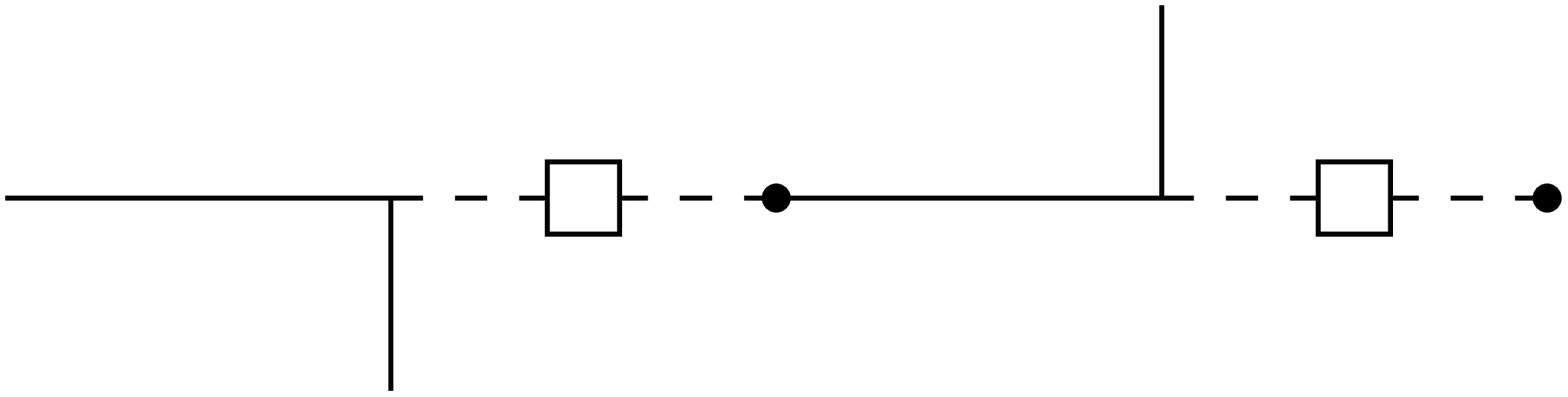,width=2.565cm,angle=-90}}
\put(0,7){\epsfig{file=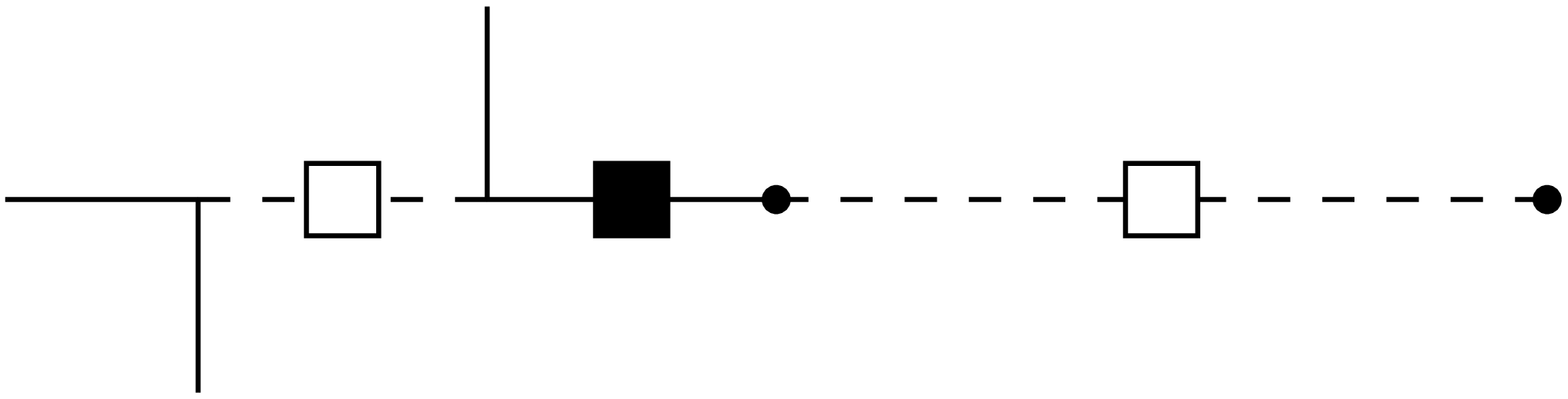,width=2.76cm,angle=-90}}
\end{picture}}
\caption{\small The diagrams for $\chi_x^{\rm{A}}(t)$ (top) and  
$\chi_x^{\rm{B}}(t)$ (bottom). The full box represents the insertion of a
 CB form factor.
The other symbols are analogous to Fig.~\ref{figureS12}. 
The upward and downward spikes symbolize the remaining 
charges.\label{figureA12}}
\end{figure}

Consider next the response function.
The contribution of group A is sketched in Fig.~\ref{figureA12}.
In the negative-time branch, the initial sojourn is followed by an
extended blip and  an extended sojourn state. Both of them
are equipped with a CS and CB form factor, respectively.
At time zero, the system hops back into 
a blip state and stays there until time $t$. The extended blip is again
decorated with a CS form factor. In mathematical terms, we have for $t>0$ 
\begin{equation}
\chi_x^{\rm{A}}(t) = \frac{2}{\hbar}\,\Delta^2\,
\sin(\epsilon t)\,e^{-\gamma t/2}_{} \int_0^\infty \! d\tau\,ds\,
\sin (\epsilon\tau)\,e^{-S(\tau)}_{}\, 
e^{-\gamma\tau/2}_{}\,e^{-\gamma s}_{} \;.
\end{equation} 
Now, as shown in Ref.~\cite{sass-w901}, the double integral times the factor
$\Delta^2$ is just
$P_{\infty}=\langle \sigma_z(t\rightarrow\infty)\rangle$. In the end, we find
\begin{align}\label{axa12t}
\chi_x^{\rm{A}}(t) &= (2/\hbar)P_{\infty}\,\sin(\epsilon t)\,e^{-\gamma t/2} 
\;,  \\
P_\infty &= \frac{2}{\pi}\,{\rm Im}\,\psi\left(\frac{1}{2} + \frac{\hbar\gamma}
{4\pi k_{\rm B}^{}T} + i \frac{\hbar\epsilon}{2\pi k_{\rm B}^{}T}\right)\;,
\end{align}
where $\psi(z)$ is Euler's digamma function. Thus we find again
exponential decay, resulting from exponential
suppression factors due to collapsed blips or 
sojourns in each interval. 

Now we turn to the contributions of group B.
It is immediately clear that the part of $\chi_x^{\rm{B}}(t)$ resulting 
from the second term in the curly bracket
of Eq.~(\ref{axb}) vanishes as $\kappa^2$ in the limit $\kappa\to 0$,
whereas the first term is nonzero in this limit. Here, the system hops
from the initial sojourn into a blip at a negative time $-\tau$ and 
stays there until time zero, where it returns to a sojourn state. At time
$s$ it hops again into a blip state and dwells in this state until time $t$.
Again, each blip interval is decorated with  a CS form factor, as discussed
above. Because of the factor $\xi_{n+1}$ in Eq.~(\ref{axb}), however, 
the extended sojourn in the positive time branch is free of collapsed
blips. The interacting dipole has length $\tau+s$ and introduces correlations 
between the negative and positive time branches (see Fig.~\ref{figureA12}). 
Thus we have
\begin{equation} \label{axb12}
\chi_x^{\rm{B}}(t)=\frac{\Delta^2}{\hbar}\int_0^\infty\! d\tau\int_0^t\! ds\,
e^{-\gamma (t+\tau - s)/2}_{}\,e^{-S(\tau+s)}_{}
\cos[\epsilon(t-\tau-s)] \;.
\end{equation}
Introducing the dipole length $\tau + s$ as a new integration variable,
performing the other integrations, and combining the resulting expression 
with Eq.~(\ref{axa12t}), we obtain
\begin{equation} \label{ax12t}
\chi_x(t)=(2/\hbar)\big[\sin(\epsilon t)F_1(t)+
\cos(\epsilon t)F_2(t)\big]
\end{equation}
with the functions
\begin{align} \label{f1}
F_1(t)&=\frac{\Delta^2}{2\gamma}\int_0^{\infty}\! d\tau\,e^{-S(\tau)}
        \sin(\epsilon\tau)
        \big( e^{-\gamma|t-\tau|/2}+e^{-\gamma(t+\tau)/2}\big)\;,\\ 
\label{f2}
F_2(t)&=\frac{\Delta^2}{2\gamma}\int_0^{\infty}\! d\tau\,e^{-S(\tau)}
        \cos(\epsilon\tau)
        \big( e^{-\gamma|t-\tau|/2}- e^{-\gamma(t+\tau)/2}\big)  \;.
\end{align}
For asymptotic times $t\gg 1/\gamma$, we find from Eq.~(\ref{ax12t}) at
zero temperature 
\begin{equation} \label{ax12long}
\chi_x(t)=\frac{8}{\pi\hbar}\frac{\gamma^2}{
\gamma^2+4\epsilon^2}\, \frac{1}{\gamma t} \;.
\end{equation}
The algebraic decay law arises from the contribution of group B.
Because of the absence of collapsed blips in the sojourn interval $s$,
this interval gets effectively very large, $s\approx t$ for $t\gg1/\gamma$.
The $1/t$ law in Eq.~(\ref{ax12long}) is simply the signature of the bare
intra-dipole interaction, $e^{-S(t)}\propto 1/t$ for $K=\frac{1}{2}$. 
The algebraic law at $T=0$ is not only of academic interest, since it is also 
valid at low but finite 
temperatures in the intermediate time regime $1/\gamma\ll t\ll\hbar\beta$.
In the asymptotic limit $t\gg \hbar\beta\gg 1/\gamma$, we find exponential 
decay,
\begin{equation}
  \chi_x(t)=\frac{16}{\hbar\gamma}\frac{\gamma^2}{
\gamma^2+4\epsilon^2}\,\frac{1}{\hbar\beta}\,e^{-\nu_1^{} t/2}_{}\;,
\end{equation}
where $\nu_1^{}=2\pi/\hbar\beta$ is the lowest bosonic Matsubara frequency.

Since we have calculated the expressions (\ref{sxa12}) and (\ref{ax12t})
independently, we are now in a position to verify whether they are consistent
with the fluctuation-dissipation theorem,
\begin{equation}\label{FDT}
 S_x(\omega)=\hbar\,\mbox{coth}\,(\hbar\beta\omega/2)\,\chi_x''(\omega)\;.
\end{equation} 
Taking the Fourier transform of $S_x(t)$ and of $\chi_x(t)$, we find
for the spectral function $S_x(\omega)$ and the absorptive part of the 
dynamical coherence susceptibility $\chi_x(\omega)$
\begin{equation}
  \label{sx12w}
\begin{split}
  S_x(\omega) 
&=\gamma\frac{\gamma^2/4+\omega^2+\epsilon^2}{(\gamma^2/4+\omega^2
+\epsilon^2)^2 -4\epsilon^2\omega^2} \;, \\ 
  \chi_x''(\omega)& =\frac{\gamma}{\hbar}\,\mbox{tanh}\,
\left(\frac{\hbar\omega}{2 k_{\rm B}^{}T}\right)
\frac{\gamma^2/4+\omega^2+\epsilon^2}{(\gamma^2/4+\omega^2+
\epsilon^2)^2-4\epsilon^2\omega^2} \;.
\end{split}
\end{equation}
This confirms that the FDT is satisfied.
Finally, the real part $\chi_x'(\omega)$ of the dynamical susceptibility
reads (we put $\epsilon=0$ for simplicity)
\begin{equation}
  \chi_x'(\omega)=\frac{8\gamma}{\pi\hbar}\frac{1}{\gamma^2+4\omega^2}\,
\mbox{Re}\,\left[
\psi\Big(\frac{1}{2}+\frac{\hbar\gamma}{4\pi k_{\rm B}^{}T}\Big) -
\psi\Big(\frac{1}{2}+i \frac{\hbar\omega}{2\pi k_{\rm B}^{}T}\Big)\right]\;.
\end{equation}
The linear static susceptibility 
$\displaystyle \chi^{(0)}_{x}= \chi'_{x}(\omega\to 0)$
diverges logarithmically as $T\to 0$.

One final remark on the case $K=\frac{1}{2}$ is appropriate. The correlation 
functions can also be calculated in a fermionic representation by
exploiting the equivalence of the 
spin-boson model for  $K=\frac{1}{2}$ with the Toulouse limit of the 
anisotropic Kondo or resonance level model~\cite{book}. One finds that the
$\sigma_x$ correlation function in the resonance level model directly
corresponds to the $\tilde\sigma_x$ correlation function of the
spin-boson model~\cite{transform}. 
The investigation of the spin-boson model is convenient when we depart from
the particular case $K=\frac{1}{2}$.

\section{The case $K=\frac{1}{2}-\kappa$}

\subsection{Expansion around $K=\frac{1}{2}$}

\begin{figure}[htb]
\setlength{\unitlength}{0.75cm}
\centerline{
\begin{picture}(9,4)
\put(0,3.8){\epsfig{file=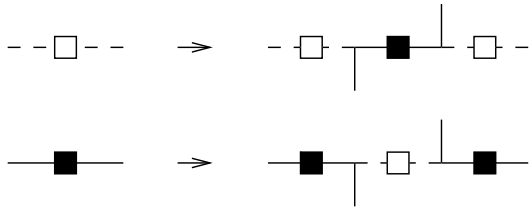,width=2.736cm,angle=-90}}
\end{picture}}
\caption{\small Insertion of extended dipoles according to rule 2. 
\label{insert}}
\end{figure}

In the previous section, we have solved the case $K=\frac{1}{2}$ in analytic
form by using the concept of collapsed blips and collapsed sojourns.
Let us now consider the regime $K=\frac{1}{2}-\kappa$ with 
$\kappa \ll 1$ and perform an expansion around the solution of the
correlation function for $K=\frac{1}{2}$.
For finite $\kappa$, a dipole is actually no longer collapsed.
The basic idea now is to develop a $\kappa$-expansion by systematically
taking into account  the finite lengths of the blips and sojourns.
To put up a general computation scheme, it is essential to split the
breathing mode integral $I(K)$ given in Eq.~(\ref{gamma}) into a contribution 
$I_1(K)$ from the short-length interval $0 <\tau<1/\tilde\gamma$
and a residual contribution $I_2(K)$ from lengths $\tau>1/\tilde\gamma$.
The inverse time scale $\tilde\gamma$ is selfconsistently determined by the
short-time part $I_1(K)$. We have 
\begin{equation}
I_1(K={\textstyle\frac{1}{2}}-\kappa)\equiv
\tilde{\gamma} = \Delta^2 \pi \kappa \int_{0}^{1/ \tilde{\gamma}} \!
d \tau \,\frac{1}{({\omega}_c \tau)^{1-2\kappa}_{}} 
\approx
\gamma \left( \frac{\omega_c}{\gamma} \right)^{2\kappa}
\; ,
\end{equation} 
where the factor $\pi\kappa$ is the remnant of the $\cos(\pi K)$ phase
factor. The frequency $\tilde\gamma$ is the effective inverse time scale
of the problem. The short-time part $I_1(K)$, representing either a
collapsed blip or a collapsed sojourn, can be treated exactly in the same 
manner as described in the previous section. That is, all possible
arrangements of collapsed dipoles within an extended sojourn and blip
of length $s$ and $\tau$, respectively,
add up to a  CB form factor $e^{-\tilde\gamma s}_{}$
and CS form factor $e^{-\tilde\gamma \tau/2}_{}$. The strategy is then
completed by developing a systematic scheme to calculate the contributions
from the extended blip and sojourn intervals, 
$\tau_j^{},\,s_j^{} > 1/\tilde\gamma$. 
Since the leading contribution $I_2(K)$ is of order $\kappa$, it is natural
to set up an expansion of the correlation function in the number of
extended dipoles. Clearly, this is not a systematic expansion in
$\kappa$ since every extended dipole also contributes higher-order
corrections in $\kappa$. However, the strategy will allow to
extract the actual long-time behavior of the correlation function. 
To perform the analysis, it is useful to introduce a diagrammatic picture.
A generic  contribution of order $\kappa^m_{}$ is obtained by adding
$m$ extended dipoles  to the respective diagram for $K=\frac{1}{2}$. Its 
structure and asymptotic behavior are essentially determined by the  
following rules which are drawn from the exact formal expressions.\\
1.) Insertion of an extended sojourn into a dressed blip interval leads
to the diagram of Fig.~\ref{insert} (top), whereas insertion of an extended
blip into a dressed sojourn interval is diagrammatically represented in
Fig.~\ref{insert} (bottom).\\
2.) An extended dipole with insertion of a CB or CS form factor has an 
effective length of order $1/\tilde\gamma$ and therefore cannot produce 
algebraic decay of the correlation function.\\
\noindent 3.) An extended dipole which is free of CB and CS form factors 
has a length 
$\tau >> 1/\tilde\gamma$. Therefore, it is sensitive to the
unscreened dipole interaction $ e^{-S(\tau)}_{} \propto\tau^{-2K}_{}$,
and its length is eventually limited by the overall length $t$.

\noindent These rules are in correspondence with the above $\xi$-rule.
We will now apply them to the symmetrized correlation function 
$S_x(t)$ and to the response function $\chi_x(t)$. 

\subsection{Response function $\chi_x(t)$}

We start the discussion of the response function by considering the
path contributions of group A, Eq.~(\ref{axa}).
In order  $\kappa^m_{}$, the diagram for $\chi{}_x^{\rm{A}}(t)$, 
Fig. \ref{figureA12} (top), is supplemented by  $m$ extended dipoles.
They can be arbitrarily distributed among the negative and positive
time branches using rule 1 (cf. Fig.~\ref{insert}). 
As a result, each interval (except for the 
first sojourn) is dressed by a CB or CS form factor.
 Thus, using argument 2, $\chi_x^{\rm{A}}(t)$ decays 
exponentially.

Consider next the contribution to $\chi_x^{\rm{B}}(t)$ from the first term in
the curly bracket of Eq.~(\ref{axb}), referred to as $\chi_x^{\rm{B}_1}(t)$.
In order $\kappa^m_{}$, the diagram in Fig.~\ref{figureA12} (bottom)
for $\chi_x^{\rm{B}}(t)$ is modified as follows. 
There are insertions of $m$ extended sojourns which can be arbitrarily 
distributed among the two blip intervals displayed [rule 1]. Each of them 
is again confined to a length of order $1/\tilde\gamma$. Due to the
factor $\xi_{n+1}$ in Eq.~(\ref{axb}), however, the
initial sojourn at positive times remains free of insertions. 
At times $t\gg 1/\tilde\gamma$, the length of this interval is therefore
 effectively $t$.
 Employing argument 3, we see that the contribution $\chi_x^{\rm{B}_1}(t)$
decays as $e^{-S(t)}_{} \propto t^{-(1-2\kappa)}_{}$. This law is
 generally valid,
only the prefactor depends on the number of extended dipoles considered.  

Starting with order $\kappa^2$, there is also a
contribution from the second term in the curly bracket of (\ref{axb}),
called $\chi_x^{\rm{B}_2}(t)$.  
The diagrams are as for $\chi_x^{\rm{B}_1}(t)$, apart from the crucial
difference that  the first sojourn in the positive time branch is dressed.
This is due to the absence of the factor $\xi_{n+1}$ 
 in $\chi_x^{\rm{B}_2}(t)$.
Thus, according to rule 2, $\chi_x^{\rm{B}_2}(t)$ decays exponentially.

\subsection{Symmetrized correlation function $S_x(t)$}

Consider first $S_x^{\rm A}(t)$, Eq.~(\ref{sxa}), which has
dynamics only in the positive time branch. Employing rule 1, we have
$m$ extended sojourns in order $\kappa^m$, and each of the blip and
sojourn intervals is dressed. Thus, $S_x^{\rm{A}}(t)$ decays 
exponentially on the time scale $1/\tilde\gamma$. 

As emphasized in the previous section, 
the leading contribution to $S{}_x^{\rm{B}}(t)$ is of order $\kappa$.
This term is found to be
\begin{equation}\label{sxeps}   
\begin{split}
{S{}_x^{\rm{B}}(t)} &= 
       -\pi \kappa \frac{\Delta^2}{2} \int_0^\infty\!d\tau\!\int_0^t ds 
\; e^{-S(\tau+s)}_{}\,e^{- \tilde{\gamma}(t+\tau-s)/2}_{}\,  
\big\{e^{- \tilde{\gamma}s}_{} -1\big\} 
     \, \cos[\,\epsilon(t-\tau -s)\,] 
\;.  
\end{split}
\end{equation}
The reason for the subtraction in the curly bracket is the missing of 
diagrams without any insertions in the sojourn interval $s$. There is 
always at least one collapsed blip due to the constraint in the 
$\xi$-summation of expression (\ref{sxb}). 
Introducing the length $\tau +s$ as a new integration variable,
the other integrations can be performed. With the definition
\begin{equation}
F_3(t) = \frac{\Delta^2}{2\tilde{\gamma}} \int_0^{\infty}\!  
       d\tau\, e^{-S(\tau)}
        \sin(\epsilon \tau) \, ( e^{- \tilde{\gamma}|t-\tau|/2} -
      e^{- \tilde{\gamma}(t+\tau)/2} )\; ,
\end{equation}
and with $\gamma$ replaced by $\tilde{\gamma}$ in the expression (\ref{f2}) 
for $F_2(t)$, we find
\begin{equation}\label{sbe}
\begin{split} 
S{}_x^{\rm{B}}(t) &=  \pi \kappa \;
    \biggl\{ \;
        \bigl[ \;  \cos(\epsilon t) F_2(t) + 
           \sin(\epsilon t) F_3(t) \; \bigr] \;   \\
&\qquad-  \frac{\Delta^2}{2} 
           \Bigl ( \; \int_{0}^{t} \! d\tau \; \tau \, e^{-S(\tau)}
                e^{- \tilde{\gamma}(\tau+t)/2} \cos[\epsilon(t-\tau)]\\ 
&\qquad\quad\;\,+ t \, \int_{t}^{\infty} \! d\tau \, e^{-S(\tau)}
                e^{- \tilde{\gamma}(\tau+t)/2} \cos[\epsilon(t-\tau)] \; 
               \Bigr)\; \biggr\} \;.
\end{split}
\end{equation}
The dominating contribution  for $t\gg 1/\tilde\gamma$ comes from the
first line in Eq.~(\ref{sbe}), yielding 
\begin{equation}
\label{sx1o}
S{}_x^{\rm{B}}(t) \, = \, 4 \kappa \;
            \frac{\tilde{\gamma}^2}{\tilde{\gamma}^2 +4 \epsilon^2}
            \; \left( \frac{1}{\tilde{\gamma} t} \right)^{1-2\kappa} 
\;.
\end{equation}
The origin of the algebraic decay for  $\kappa \neq 0$ is the subtraction 
term in the curly bracket in Eq.~(\ref{sxeps}). The analysis shows that 
the subtraction also appears 
in all higher orders in $\kappa$. Thus the asymptotic behavior 
$t^{-(1-2\kappa)}$ is generally valid for $\kappa\neq 0$. 
For $\kappa\to 0$, the prefactor of the algebraic decay law vanishes and
the decay is exponential [cf.~Eq.~(\ref{sxa12})].

\section{Long-time behavior for general $K<1$}

From the structure of the various contributions for 
$K=\frac{1}{2}-\kappa$, we can draw conclusions for the long-time
behavior of the correlation functions for general $K< 1$.
For $K$ substantially different from $\frac{1}{2}$, the modifications
concern the CB and CS form factors
inserted in a given interval. 
Instead of collapsed noninteracting blips and sojourns,
we now have extended interacting dipoles and it is no longer possible
to perform the grand-canonical sum in analytic form.  
However, according to the $\xi$-rule, the sequences of charges are
grouped into clusters which are separated by bare sojourns. 
Because of the alternating sum of the charges within a cluster, 
the length of a cluster is effectively of order $1/\Delta_r$, where 
$\Delta_r$ is the renormalized frequency  (\ref{deltared}). 

With this being the only essential modification, the asymptotic behaviors
of the  various contributions to the correlation function at times 
$t\gg 1/\Delta_r$ emerge as follows. In group A, there is a 
single neutral cluster surrounding the origin of the
time axis. Hence, both
$S_x^{\rm{A}}(t)$ and $\chi_x^{\rm{A}}(t)$ decay exponentially. In group B,
we have a charged cluster in each time branch, satisfying overall
neutrality.
Since in both branches the initial sojourn is free of insertions,
 the two clusters are near the origin and near $t$,
respectively, and they interact with the unscreened charge-charge interaction 
$e^{-S(t)} \propto t^{-2K}$ \cite{str}. This interaction directly determines 
the long-time behavior of $S_x^{\rm B}(t)$ and $\chi_x^{\rm B}(t)$.
The contributions of group B predominate over the exponential contributions
of group A for $t\gg 1/\Delta_r$. Thus we have asymptotically
\begin{align}\label{asbev1}
S_x (t) &\propto e^{-S(t)}_{} \propto t^{-2K} \;,\qquad 
K\neq{\textstyle\frac{1}{2}}\\ \label{asbev2}
\chi_x(t) &\propto e^{-S(t)}_{} \propto t^{-2K} \;.
\end{align}
Thus, the $\tilde\sigma_x$ autocorrelation function at $T=0$ decays
with a power law. The power depends on the damping strength.
Again, the $T=0$ decay laws (\ref{asbev1}) and (\ref{asbev2}) hold also at 
very low temperatures in the intermediate time regime 
$1/\Delta_r\ll t \ll \hbar\beta$. In the asymptotic limit
$t\gg\hbar\beta\gg 1/\Delta_r$, the correlation functions show exponential
decay,
\begin{equation}
  S_x(t)\propto e^{-K\nu_1 t}\;,\qquad \chi_x(t)\propto e^{-K\nu_1 t}\;,
\end{equation}
where the decay rate is $K$ times the lowest bosonic Matsubara frequency 
$\nu_1=2\pi/\hbar\beta$.
 
Let us now put the decay law (\ref{asbev1}) in perspective with the 
generalized Shiba relation
for the $\sigma_z$  correlation function 
\cite{sass-w902,chakra,costi-ki,les-sal-skor}.
In the regime $t\gg 1/\Delta_r$, this relation is expressed as
\begin{equation}\label{shiba}
S_z(t)=-2 K \,\big[\,\hbar\chi_z^{(0)}/2\,\big]^2\,\frac{1}{t^2} \;.
\end{equation}
In a charge representation for $S_z(t)$, the $1/t^2$  decay law 
reflects the dipole-dipole interaction between a
neutral cluster in the negative-time branch and a neutral cluster in 
the positive time branch. The power of the algebraic interaction is 
independent of the coupling strength and is $2$ for Ohmic dissipation.
 
Our findings are consistent with the 
fluctuation-dissipation theorem, Eq. (\ref{FDT}). 
Upon Fourier transforming Eq.~(\ref{asbev2}), we get 
$\chi''_x(\omega\to 0) \propto {\rm sgn}(\omega)|\omega|^{2K-1}$ for
$0<K<1$. Using the FDT relation (\ref{FDT}) for $T=0$, we obtain 
$S_x(\omega\to 0)$, and transforming back to 
time, we find consistency with the law (\ref{asbev1}) for $S_x(t)$.
As a by-product, we obtain a useful relation directly connecting the 
prefactors of the expressions (\ref{asbev1}) and (\ref{asbev2}),
\begin{equation}\label{pref}
   S_x(t) = (\hbar/2) \cot (\pi K)\,\chi_x(t)\;, 
\qquad t\gg 1/\Delta_r \;.
\end{equation}
In lowest order in $\kappa$, this relation is confirmed by the
results (\ref{ax12long}) and (\ref{sx1o}).
The case $K=\frac{1}{2}$ is special, since the prefactor of the $1/t$ law 
for $S_x(t)$ vanishes according to the relation
(\ref{pref}). This is in agreement with
the result (\ref{sx1o}) obtained from the direct computation of $S_x(t)$.

The asymptotic decay law (\ref{asbev2}) leads to a different behavior
for the linear (zero bias) static susceptibility 
$\displaystyle \chi_x^{(0)}=\textstyle \int_0^\infty dt\, \chi_x(t)$ 
for $K$ below and above $\frac{1}{2}$. For $K<\frac{1}{2}$, the slow decay of 
$\chi_x(t)$ implies that the linear static susceptibility diverges 
algebraically,
$\displaystyle \chi^{(0)}_x\propto T^{2K-1}$ as $T\to 0$.
This indicates that the system responds to a coherence inducing perturbation
$\propto\tilde\sigma_x$ in a nonlinear manner. Interestingly, the regime 
$K<\frac{1}{2}$ coincides with the coherence
regime for the population $\langle\sigma_z(t)\rangle$ at zero bias 
\cite{egg-grab-w,strong}. For $K>\frac{1}{2}$, the decay of $\chi_x(t)$ is
sufficiently fast so that the linear static susceptibility is finite
at $T=0$. This corresponds to the
incoherent regime for $\langle\sigma_z(t)\rangle$ at zero bias.
The transition from nonlinear to linear response at $K=\frac{1}{2}$ thus
reflects the intrinsic coherence properties of the system.
These properties of the static susceptibility have been confirmed numerically
in Ref.~\cite{strong}.

In conclusion, we have studied within a real-time approach the 
equilibrium correlation function of the
polaron-dressed tunneling or coherence operator in the dissipative 
two-state system. This quantity turns out to be universal in the scaling 
limit. The elimination of the bath modes leads to a modified influence 
functional that can be recast into the standard form at the expense of
introducing modified system paths. We have obtained the exact formal 
expressions for the coherence correlations for arbitrary damping strength $K$,
and we have presented analytic results for the particular case 
$K=\frac{1}{2}$ and for the narrow regime $K=\frac{1}{2}-\kappa$ with
$\kappa\ll 1$. 
The long-time behavior is found to be $\propto t^{-2K}$ for general $K<1$, 
reflecting the loss of coherence with increasing damping strength. 
Generally, an algebraic decay law reveals that the initial state of 
the global system is correlated. In a charge picture, the particular decay
$\propto t^{-2K}$ expresses the interaction between two clusters
with an excess charge of opposite sign. Since they 
are embedded in the vacuum ($\xi$-rule), the interaction is unscreened.

\vspace{0.3cm}
\acknowledgments
E.P. thanks R. Fazio and G. Giaquinta for useful discussions and
acknowledges financial support by the INFM under the PRA-QTMD 
programme. 
Partial support was provided by the Sonderforschungsbereich 382 of 
the Deutsche Forschungsgemeinschaft (Bonn).


\begin{references}
\bibitem{coherent} The Hamiltonian 
       $H_{\rm coh}=-\hbar\Delta\tilde\sigma_x/2+H_{\rm R,I}$, where 
      $H_{\rm R,I}$ is the second term in Eq.~(\ref{H2}), 
      effectively describes a decoupled system, 
       $\widetilde H_{\rm coh}=U^{-1}H_{\rm coh}U=
       -\hbar\Delta\sigma_x/2+H_{\rm R}$, exhibiting persistent oscillations.
\bibitem{transform} This is equivalent to calculating the autocorrelation 
                    function of the bare $\sigma_x$ with respect to the 
                    transformed Hamiltonian $\widetilde{H}=U^{-1}HU$ 
                    \cite{strong}. For $K=\frac{1}{2}$, the Hamiltonian 
                    $\widetilde{H}$ describes the Toulouse limit of the
                    resonance level model.
\bibitem{charget} In order to avoid boundary terms in the influence functional
                  (\ref{influenza}), we have included a jump at time $t$.
\bibitem{str} A similar argument was given in Ref.~\cite{strong} within an
 imaginary-time approach.
\end{references}
\end{document}